\documentclass[showpacs,twocolumn,floatfix,superscriptaddress,prb,eqsecnum]{revtex4}
\usepackage{graphicx,psfrag}
\usepackage{bm}
\usepackage{amsmath}
\begin{document}
\title{Stroboscopic model of transport through a quantum dot with spin-orbit scattering}
\author{J. H. Bardarson}
\affiliation{Instituut-Lorentz, Universiteit Leiden, P.O. Box 9506, 2300 RA
Leiden, The Netherlands}
\author{J. Tworzyd{\l}o}
\affiliation{Institute of Theoretical
Physics, Warsaw University, Ho\.{z}a 69, 00--681 Warsaw, Poland}
\author{C. W. J. Beenakker}
\affiliation{Instituut-Lorentz, Universiteit Leiden, P.O. Box 9506, 2300 RA
Leiden, The Netherlands}

\date{August 2005}
\begin{abstract}
We present an open version of the symplectic kicked rotator as a stroboscopic model of electrical conduction through an open ballistic quantum dot with
spin-orbit scattering. We demonstrate numerically and analytically that the model reproduces the
universal weak localization and weak anti-localization peak in the magnetoconductance, as predicted by random-matrix theory (RMT). We also study the transition from weak localization to
weak anti-localization with increasing strength of the spin-orbit scattering, and find agreement with RMT.
\end{abstract}
\maketitle

\section{Introduction}
Electrical conduction in semiconductor heterostructures is affected by the spin degree of freedom through spin-orbit scattering. In quantum dots with chaotic scattering a statistical approach is appropriate. The structure of the spin-orbit Hamiltonian (of either
Rashba or Dresselhaus form) has a special structure,
that of a non-Abelian vector potential. By a gauge transformation Aleiner and Fal'ko identified all possible symmetry
classes and described the crossovers between them by means of random-matrix theory (RMT).\cite{Ale01} This RMT has been extended by
Brouwer et al.\
to the case that the spin-orbit scattering is nonuniform and thus the gauge transformation cannot be
made.\cite{Bro02,Cre03} Experiments are in good agreement with the predictions of the theory.\cite{Zum02, Zum05} Recently a semiclassical
theory of quantum dots with spin-orbit scattering has been developed.\cite{Zai05a, Zai05b}

Here we present a fully quantum mechanical computer simulation to test the theory. In the case of spinless chaotic quantum dots, the
stroboscopic model known as the quantum kicked rotator
has been proven to be quite successful.\cite{Fyo00,Oss03,Jac03,Two03,Two04b, Rah05a, Rah05b} This model exploits the fact that, although the phase space of the open quantum dot is
four-dimensional, the dynamics can be described, on time scales greater than the time of flight across the dot, as a mapping between points on a two-dimensional Poincar\'e surface of section. The kicked
rotator gives a map on a two-dimensional phase space that has the same phenomenology as open quantum dots.

In this paper we extend the model of the open kicked rotator to include spin-orbit scattering in a perpendicular
magnetic field. We begin by describing the known model for a closed chaotic quantum dot\cite{Sch89} with spin-orbit scattering
in Sec.~\ref{ch:closed}, before discussing the opening up of the model in Sec.~\ref{ch:open}. The relation of the model to RMT
is given in Sec.~\ref{ch:RMT}. This relation will give us a mapping between the model parameters and the microscopic parameters of a
chaotic quantum dot. Numerical results for the weak (anti)-localization peak and its dependence on magnetic field and
spin-orbit scattering strength are presented in Sec.~\ref{ch:numerics} and compared with the analytical predictions from
Sec.~\ref{ch:RMT}.

\section{Description of the Model}
\label{ch:model}
\subsection{Closed system}
\label{ch:closed}
The symplectic kicked rotator has been introduced by Scharf~\cite{Sch89} and studied extensively in Refs.~\onlinecite{Tha93,Tha94,Oss04}. We
summarize this known model of the closed system before proceeding to the open system in the next subsection.

The symplectic kicked rotator describes an electron moving along a circle with moment of inertia $I_0$, kicked periodically at time
intervals $\tau_0$ with a kicking strength that is a function of position and spin. We choose units such that $\tau_0 \equiv 1$ and
$\hbar \equiv 1$. The
Hamiltonian $H$ is given by~\cite{Sch89,Tha93}
\begin{subequations}
  \label{eq:Hamiltonian}
\begin{align}
  H &= \frac{1}{2}(p+p_0)^2 + V(\theta)\sum_{n=-\infty}^{\infty} \delta_s(t-n), \\
  V(\theta) &=  K\cos(\theta + \theta_0) + K_\text{so}(\sigma_1\sin2\theta + \sigma_3\sin\theta).
\end{align}
\end{subequations}
We have introduced the symmetrized delta function $\delta_s(t) = [\delta(t+\epsilon) + \delta(t - \epsilon)]/2$, with $\epsilon$ an infinitesimal. 
The dimensionless angular momentum operator
$p = -i \hbar_\text{eff} \partial/\partial\theta$, with $\hbar_\text{eff}=\hbar\tau_0/I_0$ the effective Planck constant, is 
canonically conjugate to the angle $\theta \in [0,2\pi)$. The kicking potential $V(\theta)$ contains the Pauli spin matrices 
\begin{equation}
 \sigma_1 = \left(\begin{array}{cc}
0&1\\ 1&0
\end{array}\right), \quad
 \sigma_2 = \left(\begin{array}{cc}
0&-i\\ i&0
\end{array}\right), \quad
 \sigma_3 = \left(\begin{array}{cc}
1&0\\ 0&-1
\end{array}\right).
\end{equation}
Potential scattering is parameterized by the kicking strength $K$ and spin-orbit scattering by
$K_\text{so}$.

Spin rotation symmetry is broken if $K_\text{so} \neq 0$. The generalized time-reversal symmetry~\cite{Sch89} 
\begin{equation}
  \label{eq:Tsymm}
\mathcal{T}:\;\; \theta \mapsto -\theta,\;\; p \mapsto p,\;\;
\sigma_i \mapsto -\sigma_i,\;\; t \mapsto -t,
\end{equation}
is preserved if $\theta_0 = 0$ and is broken if $\theta_0 \in (0,\pi)$. A nonzero
$p_0$ ensures that the Hamiltonian has no other unitary or antiunitary symmetries.\cite{Sch89} 

Notice that the roles of $p$ and $\theta$ are interchanged in $\mathcal{T}$ compared to the conventional time-reversal symmetry of
the Rashba Hamiltonian and the spinless kicked rotator, which reads
\begin{equation}
  \label{eq:Tconv}
\mathcal{T}':\;\; \theta \mapsto \theta,\;\; p \mapsto -p,\;\;
\sigma_i \mapsto -\sigma_i,\;\; t \mapsto -t.
\end{equation}
For this reason time-reversal symmetry in the symplectic kicked rotator is broken
by a displacement of $\theta$, rather than by a displacement of $p$ as in the spinless kicked rotator.\cite{Izr90} 

The stroboscopic time evolution
of a wave function is governed by the Floquet operator 
\begin{equation}
  \mathcal{F} = \text{T}\exp\left[ -\frac{i}{\hbar_{\text{eff}}} \int_{t_0}^{t_0 + 1}H(t) dt  \right],
\end{equation}
where T denotes time ordering of the exponential. In the range $[-1/2,1/2)$ only $t_0 = 0$ and $t_0 = -1/2$ preserve
$\mathcal{T}$-symmetry for $\theta_0 = 0$. We will find it convenient to choose $t_0 = -1/2$ for numerical calculations and $t_0 = 0$
for analytical work. 

The reduction of the Floquet operator to a discrete finite form is obtained for special values of $\hbar_\text{eff}$, known as
resonances.\cite{Izr90} For $\hbar_\text{eff} = 4\pi/M$, with $M$ an integer, the Floquet operator is represented by an $M\times M$ matrix of
quaternions. (A quaternion is a $2\times 2$ matrix $q = q_0\openone + iq_1\sigma_1 + iq_2\sigma_2 + iq_3\sigma_3$ with $q_i$ a complex
number and $\openone$ the $2\times 2$ unit matrix.) For this value of $\hbar_\text{eff}$ the momentum is
restricted to $p \in [0,4\pi)$, i.e.\ one can think of the Floquet operator as describing a map on a torus. For $t_0 = -1/2$ the matrix elements in the $p$-representation are given by
\begin{subequations}
\begin{align}
\mathcal{F}_{ll'}&=(\Pi U X U^\dagger \Pi)_{ll'}, \quad l,l' = 0,1,\ldots, M-1,\\
\Pi_{ll'}&=\delta_{ll'}e^{-i\pi (l+l_0)^2/M }\openone, \quad l_0 = \frac{p_0M}{4\pi}, \\ 
U_{ll'}&=M^{-1/2}e^{-i2\pi ll'/M}\openone,\\
X_{ll'}&= \delta_{ll'}e^{-i(M/4\pi)V(2\pi l/M)}.
\end{align}
\end{subequations}
For $t_0 = 0$ one has instead 
\begin{eqnarray}
  \label{eq:Floquet0}
  \mathcal{F}= U X^{1/2} U^\dagger \Pi^2 U X^{1/2} U^\dagger.
\end{eqnarray}

The generalized time-reversal symmetry~\eqref{eq:Tsymm} is expressed by the identity
\begin{equation}
  \label{eq:DualityF}
  \mathcal{F} = \mathcal{F}^R, \quad \text{if $\theta_0 = 0$}.
\end{equation}
The superscript $R$ denotes the dual of a quaternionic matrix, 
\begin{equation*}
  \mathcal{F}^R \equiv \sigma_2\mathcal{F}^T\sigma_2.
\end{equation*}
Here $T$ denotes the transpose in the basis of eigenstates of $p$ ($p$-representation). To verify Eq.~\eqref{eq:DualityF} note that
$\sigma_2\sigma_i^T\sigma_2 = -\sigma_i$ and that the transpose in \mbox{$p$-representation} takes $\theta$ to $-\theta$.

\subsection{Open System}
\label{ch:open}
To describe electrical conduction we open up the kicked rotator, following the general scheme of Refs.~\onlinecite{Fyo00, Oss03, Jac03,Two03}.
We model a pair of $N$-mode ballistic point contacts that couple the quantum dot to electron
reservoirs, by imposing open boundary conditions 
in a subspace of Hilbert space represented by the indices $l_{k}^{(\mu)}$. The
subscript $k=1,2,\ldots N$, with $N = N_1 + N_2$, labels the modes (both spin directions),  and the superscript $\mu=1,2$ labels
the point contacts. The $N\times M$ quaternionic projection matrix
$P$ is given by
\begin{equation}
 P_{kk'} =
  \begin{cases}
    &\openone \quad \text{if } k' = l_k^{(\mu)}, \\
    &0 \quad \text{otherwise}.
  \end{cases}
\end{equation}

The matrices $P$ and $\mathcal{F}$ together determine the scattering matrix
\begin{equation}
  \label{eq:Smatrix}
S(\varepsilon)=P(e^{-i\varepsilon}-{\mathcal F}Q^TQ)^{-1}{\mathcal F}P^T,
\end{equation}
where $\varepsilon \in [0, 2\pi)$ is the quasi-energy and $Q^TQ = 1 - P^TP$. One readily verifies that $S$ is unitary.

\begin{figure}[tb]
  \begin{center}
    \psfrag{x1}{$2\pi$}
    \psfrag{x2}{$4\pi$}
    \psfrag{ z}{$0$}
    \psfrag{z }{$0$}
    \includegraphics[width=0.9\columnwidth,angle=0]{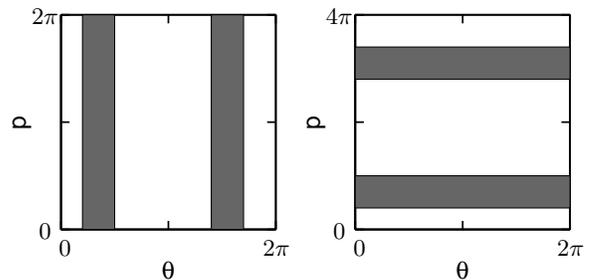}
  \end{center}
  \caption{Location of the absorbing boundary conditions (grey rectangles) in the classical phase space of the open kicked rotator. To
  ensure that the openings do not break the time reversal symmetry they are oriented parallel to the $p$-axis in the spinless kicked
  rotator (left panel) and parallel to the $\theta$-axis in the symplectic kicked rotator (right panel).}
  \label{fig:leads}
\end{figure}

We need to ensure that the introduction of the point contacts does not break the $\mathcal{T}$-symmetry 
\begin{equation}
  \label{eq:TsymmForS}
  S(\varepsilon) = S^R(\varepsilon), \quad \text{if } \theta_0 = 0,
\end{equation}
or for non-zero $\theta_0$ the more general duality relation 
\begin{equation}
  S(\theta_0) = S^R(-\theta_0).
\end{equation}
This is assured by choosing the absorbing boundary conditions in a strip parallel to the $\theta$-axis, rather than parallel to the
$p$-axis as in the spinless kicked rotator (cf.\ Fig.~\ref{fig:leads}). The difference is due to the exchange of the roles of $p$ and
$\theta$ in the time-reversal symmetry operation, compare Eqs.~\eqref{eq:Tsymm} and~\eqref{eq:Tconv}. 

By grouping together the $N_\mu$ indices belonging to the same point contact, the $N\times N$ quaternionic matrix $S$ can 
be decomposed into 4 sub-blocks containing the quaternionic transmission and reflection matrices,
\begin{equation}
S=\left(\begin{array}{cc}
r&t\\t'&r'
\end{array}\right).
\end{equation}
The value of $\varepsilon$ is arbitrary; we will take $\varepsilon = 0$ in the analytical calculations and average over $\varepsilon$ in
the numerics.
The $\mathcal{T}$-symmetry~\eqref{eq:TsymmForS} requires that $r = \sigma_2 r^T \sigma_2$, $r' = \sigma_2 {r'}\mbox{}^{ T} \sigma_2$,  and $
t' = \sigma_2 t^T \sigma_2$.

The conductance $G$ follows from the Landauer formula
\begin{equation}
\label{eq:Gtt}
G=\frac{e^2}{h}\text{Tr }\,tt^{\dagger},
\end{equation}
where the trace $\text{Tr}$ is over channel indices as well as spin indices. Unitarity of $S$ ensures that $\text{Tr } tt^\dagger = \text{Tr }
t't'\mbox{}^\dagger$. For $\theta_0 = 0$ the eigenvalues of $tt^\dagger$ are doubly
degenerate due to the $\mathcal{T}$-symmetry (Kramers degeneracy).
It will prove useful to write
the Landauer formula in the form~\cite{Bro02, Cre03}
\begin{equation}
  \label{eq:Landauer}
  G = \frac{2e^2}{h}\frac{N_1N_2}{N} - \frac{e^2}{h}{\rm Tr}S\Lambda S^{\dagger} \Lambda \equiv G_0 + \delta G,
\end{equation}
with $\Lambda$ a diagonal matrix having diagonal elements
\begin{equation}
  \Lambda_{jj} = 
  \begin{cases}
    N_2/N& j=1,\ldots, N_1, \\
    -N_1/N& j=N_1+1,\ldots, N.
  \end{cases} 
\end{equation}
The term $G_0 = (2e^2/h)N_1N_2/N$ is the classical conductance and the term $\delta G$, of order $e^2/h$, is the quantum correction from the
weak localization effect.

\section{Relation to Random-Matrix Theory}
\label{ch:RMT}
Random-matrix theory (RMT) gives universal predictions for the quantum correction $\delta G$ in Eq.~\eqref{eq:Landauer}. 
We calculate this quantity for the symplectic kicked rotator and compare with RMT. This will give us the relation between the parameters
of the stroboscopic model and the microscopic parameters of the quantum dot.

The three universality classes of RMT are labeled by $\beta = 1,2,4$, with~\cite{Bee97}
\begin{equation}
  \delta G_\text{RMT}= \frac{\beta-2}{2\beta}\frac{e^2}{h}.
\end{equation}
In the absence of $\mathcal{T}$-symmetry one has $\beta = 2$. In the presence of $\mathcal{T}$-symmetry one has $\beta = 1$ ($4$) in the
presence (absence) of spin rotation symmetry. We will investigate the three symmetry breaking transitions $\beta = 1 \rightarrow 2$, $\beta = 1
\rightarrow 4$, and $\beta = 4 \rightarrow 2$ in separate subsections.

\subsection{$\bm{\beta = 1 \rightarrow 2}$ transition}
The $\beta = 1 \rightarrow 2$ transition takes place in the absence of spin-orbit scattering ($K_\text{so} = 0$). This transition
was studied in Ref.~\onlinecite{Two04b} for the case that the symmetry $\mathcal{T}'$ rather than $\mathcal{T}$ is broken. To fully
characterize the model we need to reconsider this transition for the case of $\mathcal{T}$-symmetry breaking.

For small $\theta_0$, $\cos(\theta + \theta_0) \approx \cos\theta - \theta_0 \sin\theta$ and the Floquet matrix~\eqref{eq:Floquet0} takes the form 
\begin{subequations}
\begin{align}
  \label{eq:Fbeta1to2}
  &\mathcal{F}(K_\text{so} =0, \theta_0\rightarrow 0) = e^{\theta_0W}\mathcal{F}_0e^{\theta_0W}, \\
  &W = UYU^\dagger, \quad Y_{ll'} = \delta_{ll'}i(KM/8\pi)\sin (2\pi l/M).
\end{align}
\end{subequations}
Here $\mathcal{F}_0 = \mathcal{F}(K_\text{so}=0,\theta_0=0)$ is
unitary symmetric and $W$ is real antisymmetric. The scattering matrix~\eqref{eq:Smatrix} (at $\varepsilon = 0$) becomes
\begin{subequations}
\begin{align}
  \label{eq:Su}
  S &= T(1-\mathcal{F}_0R)^{-1}\mathcal{F}_0T',\\
  T &= Pe^{\theta_0W},\\
  T' &= e^{\theta_0W}P^T, \\ 
  R &= e^{\theta_0W}Q^{T}Qe^{\theta_0W}. 
\end{align}
\end{subequations}
Substitution of $S$ into Eq.~\eqref{eq:Landauer} gives the conductance $G$.

To make contact with RMT we assume that $\mathcal{F}_0$ is a random matrix from the circular orthogonal ensemble (COE), expand the expression for $G$ in powers of
$\mathcal{F}_0$ and average $\mathcal{F}_0$ over the COE. In the regime $1 \ll N \ll M$, we can perform the average over the unitary
group with the help of the diagrammatic technique of Ref.~\onlinecite{Bro96}. Since
$\text{Tr } \Lambda = 0$ only the maximally crossed diagrams contribute to leading order in $N$. The result for the average quantum
correction becomes
\begin{equation}
  \label{eq:avG}
  \langle \delta G \rangle = -\frac{2e^2}{h}\text{tr } T^\dagger\Lambda T (T'\Lambda T'^\dagger)^T \frac{1}{M - \text{tr } R^\dagger R^T}.
\end{equation}
The factor of $2$ comes from the spin degeneracy and the trace tr is over channel indices only.
The two remaining traces are evaluated in the limit $N,M \rightarrow \infty$ at fixed $N/M$. We find 
\begin{align}
  &M^{-1}\text{tr } T^\dagger\Lambda T (T'\Lambda T'^\dagger)^T = \frac{N_1N_2}{N^2}\frac{N}{M}, \\
  &M^{-1}\text{tr } R^\dagger R^T = 1 - N/M - \theta_0^2(KM/4\pi)^2(1-N/M).
\end{align}
Substitution into Eq.~\eqref{eq:avG} gives the average quantum correction
\begin{subequations}
  \label{eq:wl}
\begin{align}
  \langle \delta G \rangle &= -\frac{e^2}{h}\frac{2N_1N_2}{N^2}\frac{1}{1 + (\theta_0/\theta_c)^2}, \\
  \label{eq:theta_c}
  \theta_c &= \frac{4\pi \sqrt{N}}{KM^{3/2}}.
\end{align}
\end{subequations}

The RMT result has the same Lorentzian profile~\cite{Plu94,Bee97}
\begin{subequations}
  \label{eq:wlRMT}
\begin{align}
 \delta G_\text{RMT} &= -\frac{e^2}{h}\frac{2N_1N_2}{N^2}\frac{1}{1 + (B/B_c)^2}, \\
 B_c &= C\frac{h}{eL^2}\left(\frac{NL\Delta}{\hbar v_F}\right)^{1/2},
\end{align}
\end{subequations}
with $C$ a numerical constant of order unity, $L$ the size of the dot, $\Delta$ its mean level spacing and $v_F$ the Fermi velocity. Comparison of Eqs.~\eqref{eq:wl} and~\eqref{eq:wlRMT} allows us to identify
\begin{equation}
  \theta_0/\theta_c = B/B_c.
\end{equation}

\subsection{$\bm{\beta = 1 \rightarrow 4}$ transition}
The $\beta = 1 \rightarrow 4$ transition is realized by turning on spin-orbit scattering ($K_\text{so}$) in the absence of a
magnetic field ($\theta_0 = 0$). In this transition the quaternionic structure of the Floquet matrix plays a role.
The Floquet matrix~\eqref{eq:Floquet0} has the form
\begin{subequations}
\begin{align}
  &\mathcal{F}(K_\text{so}, \theta_0 = 0) = e^{K_\text{so}A}\mathcal{F}_0e^{K_\text{so}A}, \\
  &A = U(\sigma_1Y_1 + \sigma_3Y_3)U^\dagger, \\
  &(Y_1)_{ll'} = -\delta_{ll'}i (M/8\pi) \sin (4\pi l/M), \\ 
  &(Y_3)_{ll'} = -\delta_{ll'}i (M/8\pi) \sin (2\pi l/M).
  \label{eq:A}
\end{align}
\end{subequations}
The matrix $A$ is real antisymmetric and thus $A^* = -A$, where ${}^*$ denotes quaternion complex conjugation. (The complex
conjugate of a quaternion $q$ is defined as $q^* = q_0^*\openone + iq_1^*\sigma_1 + iq_2^*\sigma_2 + iq_3^*\sigma_3$.) The scattering matrix takes the
same form~\eqref{eq:Su}, but now
with
\begin{subequations}
\begin{align}
T &= Pe^{K_\text{so}A},\\
T' &= e^{K_\text{so}A}P^T, \\ 
R &= e^{K_\text{so}A}Q^{T}Qe^{K_\text{so}A}. 
\end{align}
\end{subequations}

The average of $\mathcal{F}_0$ over the ensemble of unitary symmetric matrices only
involves the channel indices and not the spin indices. To keep the quaternions in the correct order 
we adopt the tensor product notation of Brouwer et al.\cite{Bro02,Cre03}
The average of $\delta G$ over $\mathcal{F}_0$ gives, to leading order in $N$,
\begin{equation}
  \label{eq:DeltaG1to4}
  \langle \delta G \rangle = \frac{e^2}{h}\sum_{\mu\nu} \left[ \tau \frac{\text{tr }E\otimes E'^*}{M\openone\otimes\openone - \text{tr
  }R\otimes R^*}\tau\right]_{\mu\nu;\mu\nu},
\end{equation}
where $\tau = 1\otimes\sigma_2$, $E = T^{\dagger}\Lambda T$, and $E' = T'\Lambda T'^\dagger$. 
The tensor product has a backward
multiplication in the second argument,
\begin{equation}
  (a\otimes b)(c \otimes d) \equiv ac\otimes db,
\end{equation}
and the indices $\mu$ and $\nu$ are the spin indices. 

The two traces are calculated in the limit $K_\text{so} \rightarrow 0$, $N,M \rightarrow \infty$ at fixed $N/M$, leading to
\begin{subequations}
\begin{align}
  &M^{-1}\text{tr } E\otimes E'^* = \frac{N_1N_2}{N^2}\frac{N}{M} \openone\otimes\openone, \\
  &M^{-1}\text{tr }R\otimes R^* = (1 - N/M)(1 - 4K_\text{so}^2(M/8\pi)^2)\openone\otimes\openone \nonumber\\
  &\quad\mbox{+ } 2K_\text{so}^2(M/8\pi)^2(1-N/M)(\sigma_1\otimes\sigma_1 + \sigma_3\otimes\sigma_3). 
\end{align}
\end{subequations}
After substitution into Eq.~\eqref{eq:DeltaG1to4} there remains a matrix structure that can be inverted, resulting in
\begin{subequations}
  \label{eq:wlwal}
\begin{gather}
  \langle \delta G \rangle = \frac{e^2}{h} \frac{N_1N_2}{N^2}\left( 1 - \frac{2}{1 + 2a^2} -
  \frac{1}{1 + 4a^2} \right), \\
  \label{eq:K_c}
  a = K_\text{so}/K_c, \quad K_c = \frac{4\pi\sqrt{2N}}{M^{3/2}}.
\end{gather}
\end{subequations}
The RMT result has the same functional form~\cite{Bro02}, with $a = (2\pi\hbar N/\tau_{\text{so}}\Delta)^{1/2}$.
Here $\tau_{\text{so}}$ is the spin-orbit scattering time. Thus we identify
\begin{equation}
  K_\text{so}/K_c = (2\pi\hbar N/\tau_{\text{so}}\Delta)^{1/2}.
\end{equation}

\begin{figure}[b]
  \begin{center}
    \hspace{2.6mm}\includegraphics[width=0.60\columnwidth,angle=270]{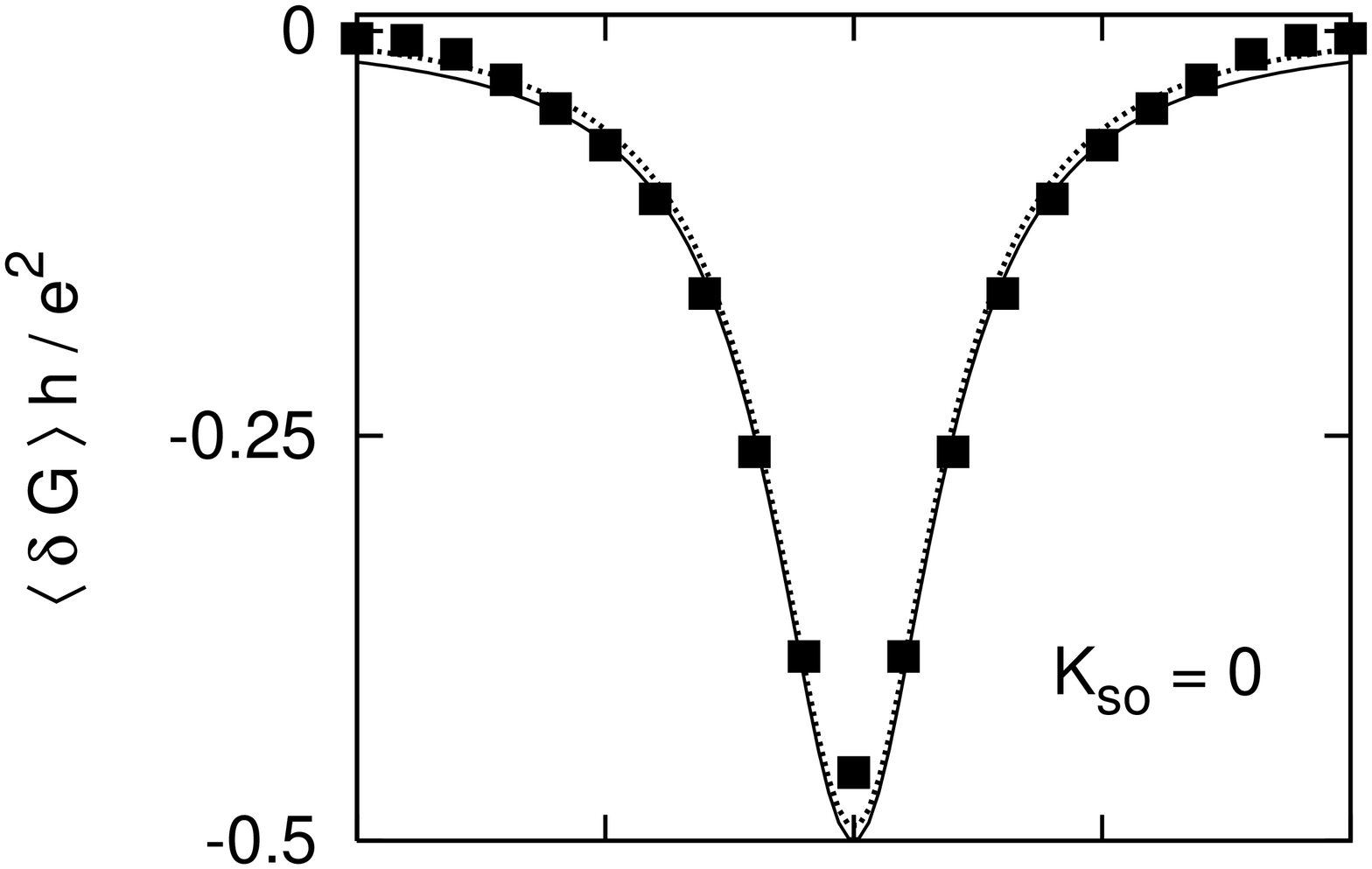}\vspace{-0.5cm} 
    \includegraphics[width=0.628\columnwidth,angle=270]{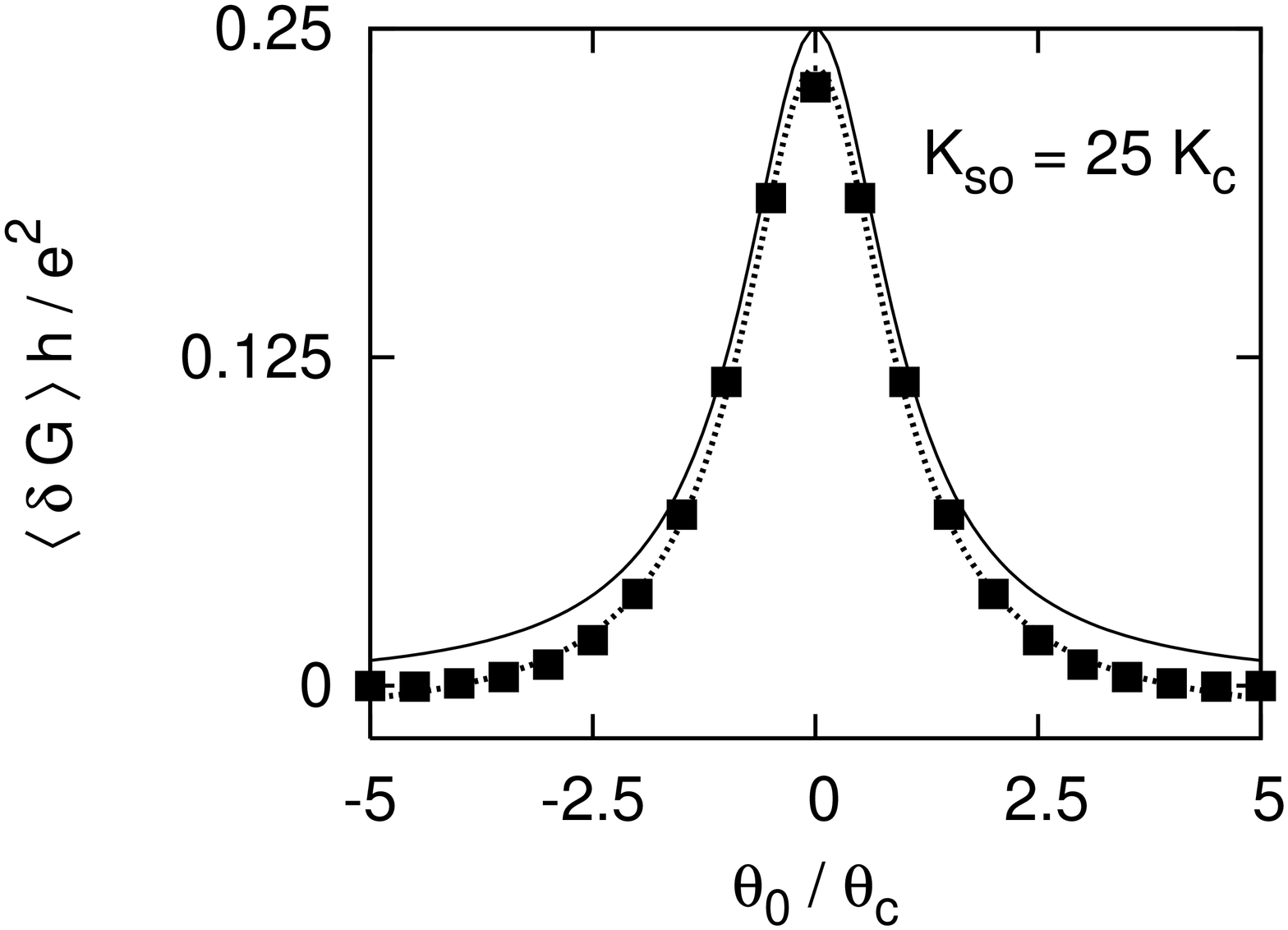}
  \end{center}
  \caption{Average quantum correction $\langle \delta G \rangle$ to the conductance as a function of the $\mathcal{T}$-symmetry breaking parameter
  $\theta_0$. The data points are for the symplectic kicked rotator characterized by $K = 41$, $M = 500$, $N = 10$, $l_0 = 0.2$. The solid lines are the analytical
  predictions~\eqref{eq:wl} and~\eqref{eq:wal} in the absence and presence of spin-orbit scattering. The dotted lines are the solid lines
  with a vertical offset, to account for a difference between the predicted and actual value of the classical conductance $G_0$.}
  \label{fig:wlANDwal}
\end{figure}

\subsection{$\bm{\beta = 4 \rightarrow 2}$ transition}
In the presence of strong spin-orbit scattering ($K_\text{so} \gg K_c$) the Floquet  matrix takes for small $\theta_0$ the same form as in
Eq.~\eqref{eq:Fbeta1to2} for $K_\text{so}=0$, but now $\mathcal{F}_0 = \mathcal{F}(K_\text{so} \gg K_c, \theta_0 = 0)$ is a unitary
self-dual matrix rather than a unitary symmetric matrix. We can repeat exactly the same steps as we
did for $K_\text{so}=0$ but with $\mathcal{F}_0$ a random matrix in the circular symplectic ensemble (CSE). We then average
$\mathcal{F}_0$ over the CSE. This leads to 
\begin{equation}
  \label{eq:wal}
  \langle \delta G \rangle  = \frac{e^2}{h}\frac{N_1N_2}{N^2}\frac{1}{1 + (\theta_0/\theta_c)^2},
\end{equation}
with $\theta_c$ as in Eq.~\eqref{eq:theta_c}. The width of the Lorentzian is therefore the same in the $\beta = 1 \rightarrow 2$ and
$\beta = 4 \rightarrow 2$ transitions, in agreement with RMT.\cite{Bee97}
\section{Numerical results}
\label{ch:numerics}
\begin{figure}[tb]
  \begin{center}
    \includegraphics[width=0.6\columnwidth,angle=270]{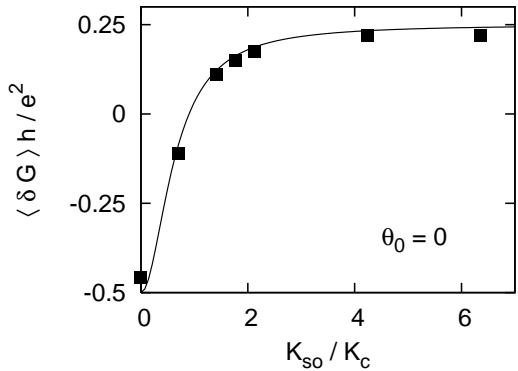}
  \end{center}
  \caption{Average quantum correction $\langle \delta G \rangle$ to the conductance as a function of spin-orbit scattering strength $K_\text{so}$ at zero
  magnetic field. Other parameters are the same as in Fig.~\ref{fig:wlANDwal}. The solid line is the analytical
  prediction~\eqref{eq:wlwal} for the crossover from weak localization to weak anti-localization.}
  \label{fig:wlwal}
\end{figure}

The numerical technique we use is the same as has been used before for the spinless kicked
rotator.\cite{Two03,Two04b} A combination of an iterative procedure for matrix inversion and the fast-Fourier-transform algorithm allows
for an efficient
calculation of the scattering matrix from the Floquet matrix.

The average conductance $\langle G \rangle$ was calculated with the Landauer formula~\eqref{eq:Gtt} by averaging over 60 different uniformly distributed quasi-energies and 40 randomly chosen lead
positions. The quantum correction $\langle \delta G \rangle$ is obtained by subtracting the classical conductance $G_0$. The numerical data is
shown in Figs.~\ref{fig:wlANDwal} and~\ref{fig:wlwal}. The magnetic field parameter $\theta_0$ is given in units of $\theta_c$ from
Eq.~\eqref{eq:theta_c} and the spin-orbit scattering strength parameter $K_\text{so}$ is given in units of $K_c$ from Eq.~\eqref{eq:K_c}. The solid lines are the analytical
predictions~\eqref{eq:wl}, \eqref{eq:wlwal}, and~\eqref{eq:wal} without any fitting parameter.

The small difference between the data and the predictions can be attributed to an uncertainty in the value $G_0$ of the classical
conductance. A small vertical offset (corresponding to a change in $G_0$ of about $0.1 \%$) can correct for this (dotted lines in
Fig.~\ref{fig:wlANDwal}). The strongly non-Lorentzian lineshape seen by Rahav and Brouwer\cite{Rah05a, Rah05b} in the spinless kicked rotator is
not observed here.

\section{Conclusion}
We have presented a numerically highly efficient model of transport through a chaotic ballistic quantum dot with spin-orbit scattering,
extending the earlier work on the spinless kicked rotator.
Through a simple assumption of a random Floquet matrix we have derived analytical predictions for the conductance of the model as a
function of spin-orbit scattering strength and magnetic field. The functional form of the conductance coincides with random-matrix theory and through this
correspondence we obtain a mapping from microscopic parameters to model parameters. Numerical calculations are in good agreement with the
analytical predictions. This model may provide a starting point for the study of the effect of a finite Ehrenfest time on weak
anti-localization, which requires very large system sizes (cf.\ Refs.~\onlinecite{Rah05a} and~\onlinecite{Rah05b}).

\section*{ACKNOWLEDGMENTS}
This work was supported by the Dutch Science Foundation NWO/FOM. We acknowledge support by the European Community's Marie Curie
Research
Training Network under contract MRTN-CT-2003-504574, Fundamentals of Nanoelectronics.

\end{document}